\documentclass[elsart3p,letter]{elsart}

 \usepackage{graphicx}
 \usepackage{subfigure}
 \usepackage{epstopdf}

\usepackage{amssymb}

\pdfoutput=1

 \usepackage{lineno}

\begin{document}

\begin{frontmatter}



\title{A Measurement of Photon Production in Electron Avalanches in CF$_4$}

\author[MIT]{A.~Kaboth\corauthref{cor}}
\corauth[cor]{Corresponding author.} 
\ead{akaboth@mit.edu}
\author[MIT]{J.~Monroe}
\author[BU]{S.~Ahlen}
\author[MIT]{D.~Dujmic}
\author[MIT]{S.~Henderson}
\author[MIT]{G.~Kohse}
\author[MIT]{R.~Lanza}
\author[BU]{M.~Lewandowska}
\author[BU]{A.~Roccaro}
\author[MIT]{G.~Sciolla}
\author[Brand]{N.~Skvorodnev}
\author[BU]{H.~Tomita}
\author[MIT]{R.~Vanderspek}
\author[Brand]{H.~Wellenstein}
\author[MIT]{R.~Yamamoto}
\author[MIT]{P.~Fisher}

\address[BU]{Boston University, Boston, MA 02215}
\address[Brand]{Brandeis University,  Waltham, MA 02454}
\address[MIT]{Massachusetts Institute of Technology, Cambridge, MA 02139}

\begin{abstract}
This paper presents a measurement of the ratio of photon to electron production and the scintillation spectrum in a popular gas for time projection chambers, carbon tetrafluoride (CF$_4$), over the range of 200 to 800 nm; the ratio is measured to be 0.34$\pm$0.04. This result is of particular importance for a new generation of dark matter time projection chambers  with directional sensitivity which use CF4 as a fill gas.
\end{abstract}

\begin{keyword}
Gas properties \sep Gas scintillation \sep Optical readout

\end{keyword}
\end{frontmatter}

\section{Introduction}

An interesting unsolved problem in physics is the nature of dark matter. Astronomical observations have shown that dark matter comprises approximately 22\% of the energy in the universe~\cite{spergel2003wc}, yet there is no strong evidence for direct detection of a dark matter particle. One of the most popular dark matter candidates is the lightest supersymmetric particle, the neutralino~\cite{ellis2005udd}. The motion of the solar system about the galactic center is expected to produce an apparent dark matter wind~\cite{drukier1986dcd}. Directional dark matter experiments use gas-filled time projection chambers with electronic~\cite{alner2005did,miuchi2007dsd} or optical readout~\cite{dujmic2007oht} to search for this wind. The signal is a directional asymmetry in dark matter induced nuclear recoils. Recoiling nuclei are detected via energy deposition in the gas, producing scintillation photons and ionization electrons.

 A particularly appealing gas is carbon tetrafluoride, CF$_4$, because supersymmetric dark matter has an enhanced spin-dependent cross-section with fluorine~\cite{tovey2000nmi}.  CF$_4$ gas also has a number of experimentally desirable features: it has a high electron drift velocity, typically 10 cm/$\mu$s~\cite{vavra1992wcg}, and it emits a large number of scintillation photons in the UV and visible light regions of the spectrum~\cite{pansky1995sca}, with a significant fraction of the photons falling in the visible region of the spectrum. This last feature is critical for experiments aiming to employ optical readout of detectors with CCD cameras~\cite{dujmic2007oht}.

This paper presents a measurement of the ratio of photons to electrons produced in CF$_4$ over the range 200-800 nm. Panksy, et al.~\cite{pansky1995sca} measured this ratio to be 0.3$\pm$0.15 in the wavelength range of 160-600 nm.  Extending the range to 800 nm is interesting because many CCDs have high quantum efficiencies in the 600-800 nm range. This paper also presents the first wavelength dependent measurement of the scintillation spectrum of CF$_4$ between 200 and 800 nm.

\section{Experimental Apparatus}

A single wire proportional tube produces both electrons and photons from an electron-induced avalanche. The proportional tube is supplied with a Fe-55 source, which primarily produces K$_{\alpha}$ X-rays at 5.89~keV~\cite{lide1995hca}. Since the work function of CF$_4$ is 54~eV~\cite{sharma1998slac}, this results in approximately 110 primary electrons in the tube.  These electrons are accelerated by a high electric field and collide with gas molecules, ionizing or exciting the target molecule in the collision. The electrons liberated by the ionization are in turn accelerated and ionize more molecules. This process creates an avalanche, which creates an electrical signal on the central axis wire, in conjunction with copious associated photons. Typical gains in this apparatus are of order $10^5$.

The proportional chamber used for this measurement is a 6.5" long, 1" inner diameter copper tube with a 50~$\mu$m diameter copper wire in the center. The tube is operated at voltages on the central wire ranging from 2250-2425~V, with the tube walls at ground, which gives rise to electric fields ranging from 30-35 kV/m near the tube walls and 8100-8800~kV/m near the wire. The tube sits in a vacuum vessel with a 1" thick acrylic top. CF$_4$ gas is supplied to this vessel, such that the gas fills both in and around the proportional tube. Typical operating pressures range from 140-180~Torr at 24-27$^{\circ}$C. 

The electron signal is collected from the central wire of the proportional tube. High voltage is supplied to the proportional tube through a 1000~k$\Omega$ resistor, and the signal is read out through a 1~nF blocking capacitor and a 1.5~k$\Omega$ resistor. The signal then is processed by an integrating preamplifier and an ORTEC model 672 spectroscopic amplifier. The output of the spectroscopic amplifier is fed into a LeCroy WaveSurfer 432 oscilloscope which acts as a trigger and analog-to-digital converter. 

The photon signal is collected with a photomultiplier tube (PMT) placed on top of the acrylic vacuum vessel cover and centered over a 1" long and 1/16" wide slit, which is covered with a 1~mm thick quartz window, in the top of the proportional tube. The entire vacuum vessel and PMT combination is placed inside a dark box. The PMT used is a Hamamatsu H1161~\cite{hamam_photot_h1161}. The output of the PMT is fed into an ORTEC model 575A integrating preamplifier and a Canberra model 2005 spectroscopic amplifier and then into the oscilloscope. 

\begin{figure}[htbp]
\centering
\subfigure[]{\label{fig:box}\includegraphics[width=0.9\linewidth]{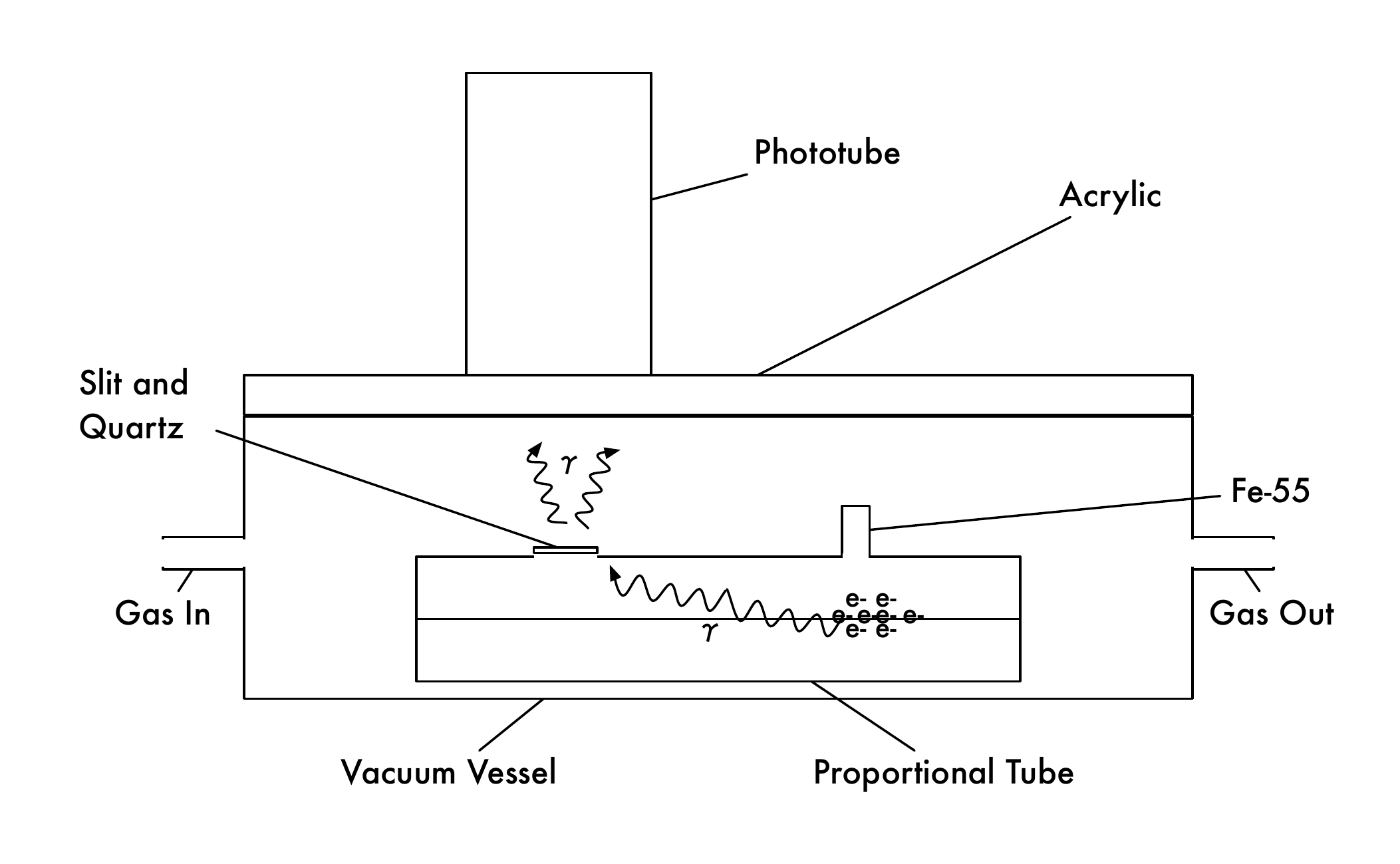}}
\subfigure[]{\label{fig:electronics}\includegraphics[height=0.85\linewidth]{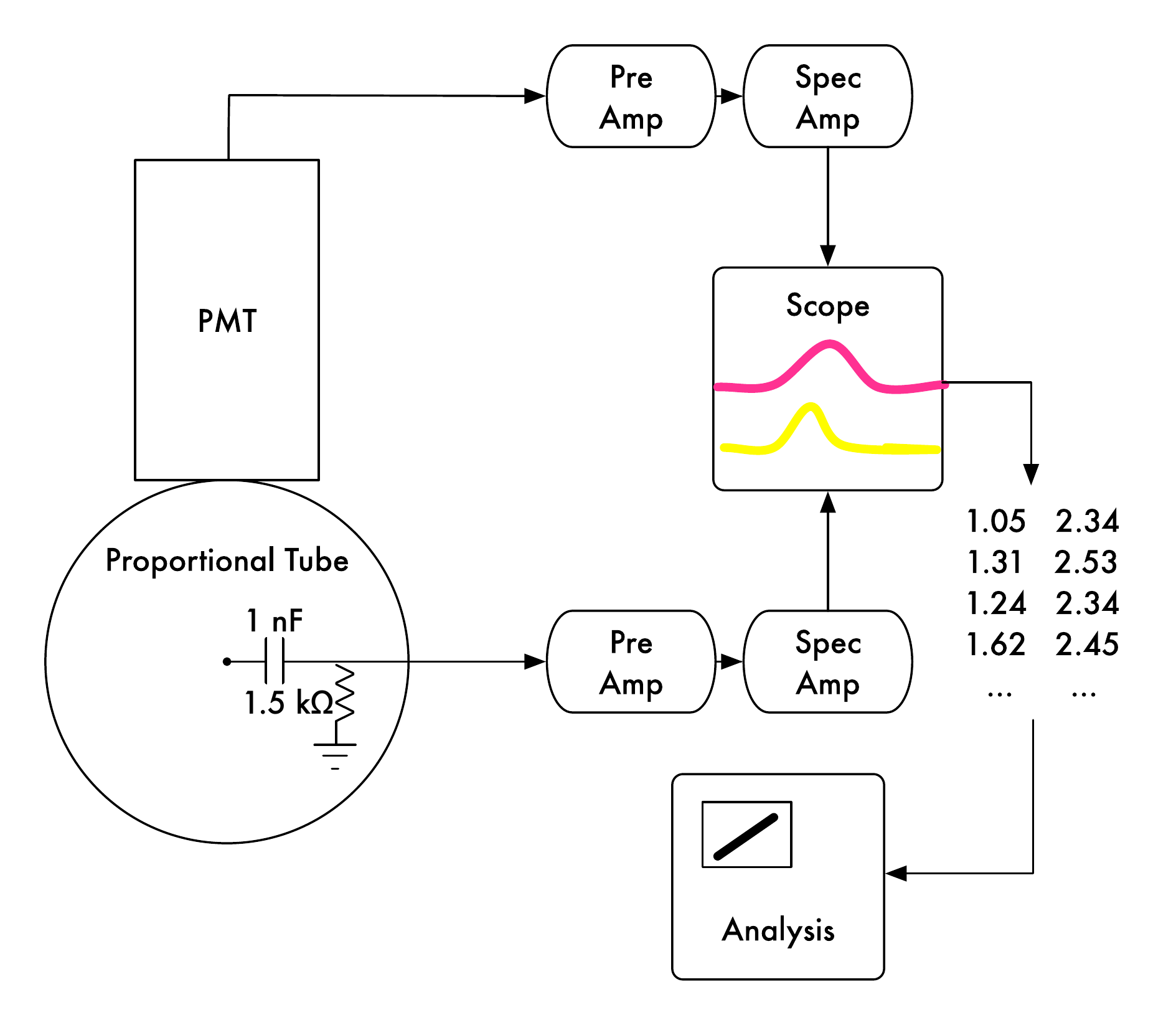}}
\caption{(a): Side view, not to scale, of experimental apparatus. (b): Schematic of the electron and photon data acquisition chain.}
\label{fig:setup}
\end{figure}

Fig.~\ref{fig:box} shows a sketch of the apparatus. Fig.~\ref{fig:electronics} shows a schematic of the electronics chain. Table~\ref{tab:electronics} shows typical signal sizes through the apparatus at a pressure of 180~Torr and 2375~V.

The purpose of this apparatus is to measure the photon to electron ratio in CF$_4$ in a generally applicable way. Thus, the efficiency and the acceptance of the apparatus is taken into account through a series of corrections described in the next section. These corrections include geometric acceptance, wavelength dependent transmission factors, PMT quantum efficiency, and the shape of the CF$_4$ photon emission spectrum.

\begin{table*}
   \caption{Typical signal sizes through the data acquisition chain.}
   \begin{tabular}{|c|c|c|} 
   	\hline
	Portion of Electronics&Size(mV)&Duration or frequency ($\mu$s)\\
       	\hline
	Wire&5-15&3-6 pulses every 0.105\\
	PMT &50-200&3-6 pulses every 0.105\\
	Wire preamplifier&50-300&5\\
	PMT preamplifier (incl. 10x atten.)&15-75&50\\
	Wire spectroscopic amplifier&500-2000&5\\
	PMT spectroscopic amplifier&120-750&15\\
	\hline
	PMT with 1 photoelectron (p.e.)&30-50&20ns\\
	PMT with 1 p.e. preamplifier (incl. 10x atten.)&1-2&50\\
	PMT with 1 p.e. spectroscopic amplifier&6-10&15\\
	\hline
   \end{tabular}

   \label{tab:electronics}
\end{table*}

\section{Calibrations}

\subsection{Electron Signal}
To measure the number of electrons in the avalanche, the amplitude of the wire signal spectroscopic amplifier pulses must be correlated to an absolute number of electrons. This is accomplished using a 10~pF test input on the integrating preamplifier. A known voltage placed across this capacitor translates into a known charge propagated through the electronics chain. A quadratic polynomial fits the calibration data best, corresponding to a small non-linearity in the preamplifier; a typical calibration is $N_{e^-} = (2.68\times~10^{6})V_{sa}^2+(4.50\times~10^7 )V_{sa}+(4.62\times~10^6)$, where $N_{e^-}$ is the number of electrons and $V_{sa}$ is the spectroscopic amplifier output. Thus, a 4~V signal in the amplifier corresponds to 2.28$\times 10^8$ electrons.

\subsection{PMT Signal}
The number of photons is measured with a PMT, which is calibrated with a known light source. For this calibration, a green (565~nm) LED is placed in the apparatus next to the slit in the proportional chamber. The LED is pulsed with a narrow (100~ns) square wave at 1~kHz. The voltage of the square wave is adjusted until a single photon from the LED is detected by the PMT in about 10\% of pulses, and zero photons the remainder of the time. Poisson statistics dictate that at this 10\% occupancy, a two photon signal would be observed 0.5\% of the time. This one photoelectron signal is fed through the PMT electronics chain and read out in the same way as a photon signal from an avalanche. The distribution of the number of events vs. PMT signal voltage is fit with two gaussians, corresponding to a pedestal and a signal peak. The mean of the signal peak corresponds to one photoelectron, and typically is 5 mV after pedestal subtraction. This calibration was repeated with 5\% and 1\% occupancy, and the results were consistent with the calibration at 10\% occupancy. Thus, a 20~mV PMT signal corresponds to 4 photoelectrons.

\subsection{Solid Angle Acceptance}
The slit in the proportional tube only captures a small fraction of the isotropic light that is emitted in the avalanche at the approximate location of the interaction point of the Fe-55 X-rays. To measure this effect, the LED is attached to an optical fiber, which is placed in the opening for the Fe-55 source. The length of the fiber is adjusted  so that the end of the fiber is as close to the central wire as possible, since the bulk of the avalanche occurs within one wire radius of the wire's surface. The PMT is placed as shown in Fig.~\ref{fig:box}, and signals are read out through the entire electronics chain.  Then the end of the fiber is affixed directly to the face of the PMT and again the signals are read out. The ratio of the mean PMT signal distribution from the fiber inside the tube to the fiber on the PMT gives a solid angle and transmission coefficient of $0.00073 \pm 0.00012$. The error on this number reflects the variation over several trials.

However, this value must be slightly corrected since the optical fiber is not isotropic, but rather directs most of the light in one direction. This correction is done using a simulation which bounces photons off of the reflective surface of a cylinder. The difference between light tightly focused in one direction and isotropic light gives rise to a correction of +5\%.  Dividing by the value of acrylic transmission at 565~nm (see Section~\ref{sec:transmission}) gives the value of the solid angle acceptance alone. No correction is made for the transmission of the quartz window, as the window had been removed when this measurement was made. Thus, the final value for the solid angle acceptance is $0.00083 \pm 0.00017$, where the error includes the measurement error, along with a 5\% error related to the anisotropy and a small error from the transmission correction. A calculation of just the angle subtended by the slit gives a solid angle factor of 0.0002. This calculation is confirmed by setting the reflectivity of the simulation to zero.

\subsection{Detector Transmission Functions}
\label{sec:transmission}

Light from the avalanche is attenuated by the quartz window and the acrylic vacuum vessel cover before reaching the PMT. The transmission of both materials is measured with a Jobin-Yvon 1250M spectrometer using an ultraviolet-sensitive Hamamatsu R928 multialkali PMT~\cite{hamam_photot_r928}, with an incandescent lightbulb providing a continuous input spectrum.

\begin{figure}[htbp] 
   \centering
   \includegraphics[height=0.8\linewidth]{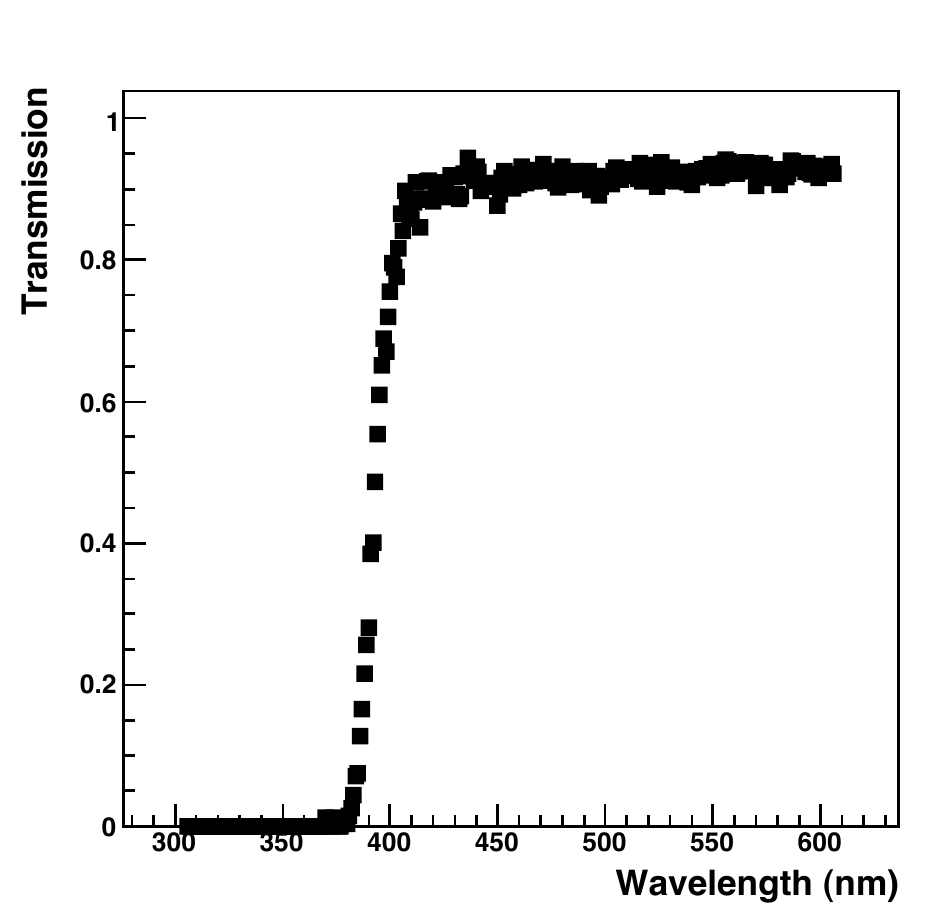} 
   \caption{Measured fractional transmission through acrylic, using an incandescent light source, versus wavelength in nm.}
   \label{fig:lucite}
\end{figure}

The measured transmission curve of the acrylic is shown in Fig.~\ref{fig:lucite}. 

\begin{figure}[htbp] 
   \centering
   \includegraphics[height=0.8\linewidth]{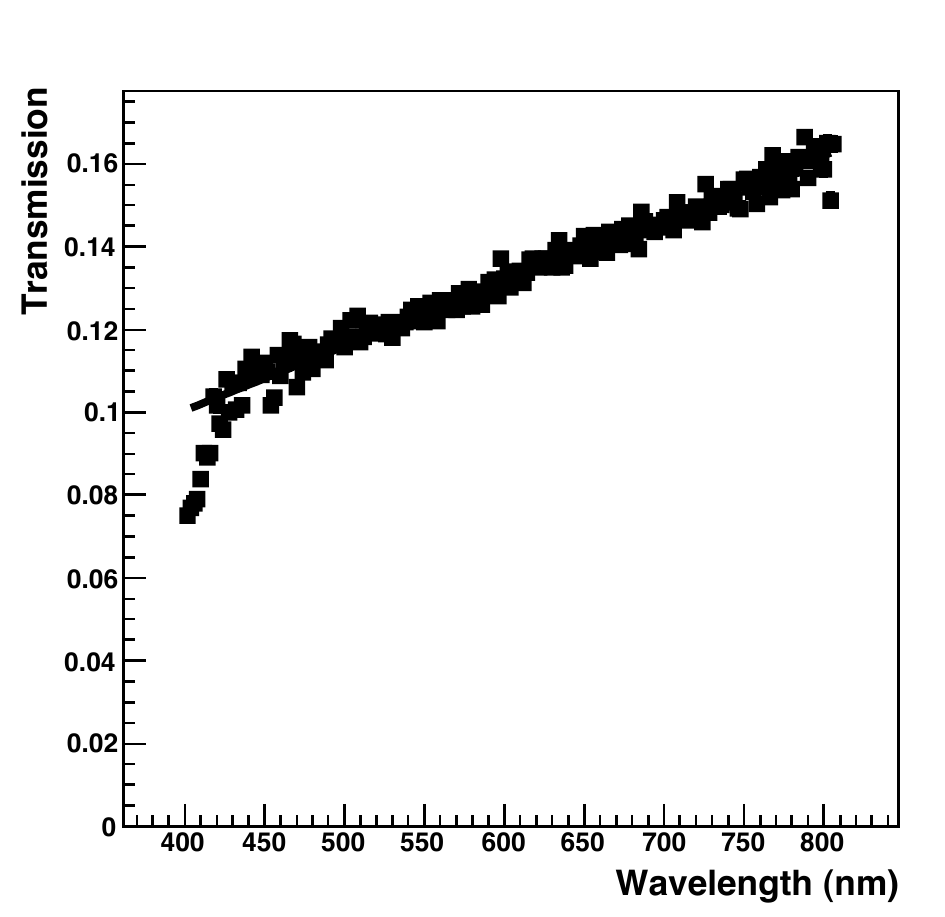} 
   \caption{Measured fractional transmission through quartz, using an incandescent light source, versus wavelength in nm.}
   \label{fig:quartz}
\end{figure}

The measured quartz transmission curve is shown in Fig.~\ref{fig:quartz}. This result is startling because the transmission is so low; the normal transmission of quartz is around 0.95 across all wavelengths of interest. This piece of quartz, however, had a crystalline growth on it, which accounts for the severely attenuated transmission. The level of growth was assumed to be constant over one month of data collection.

\begin{figure}[htbp] 
   \centering
   \includegraphics[height=0.8\linewidth]{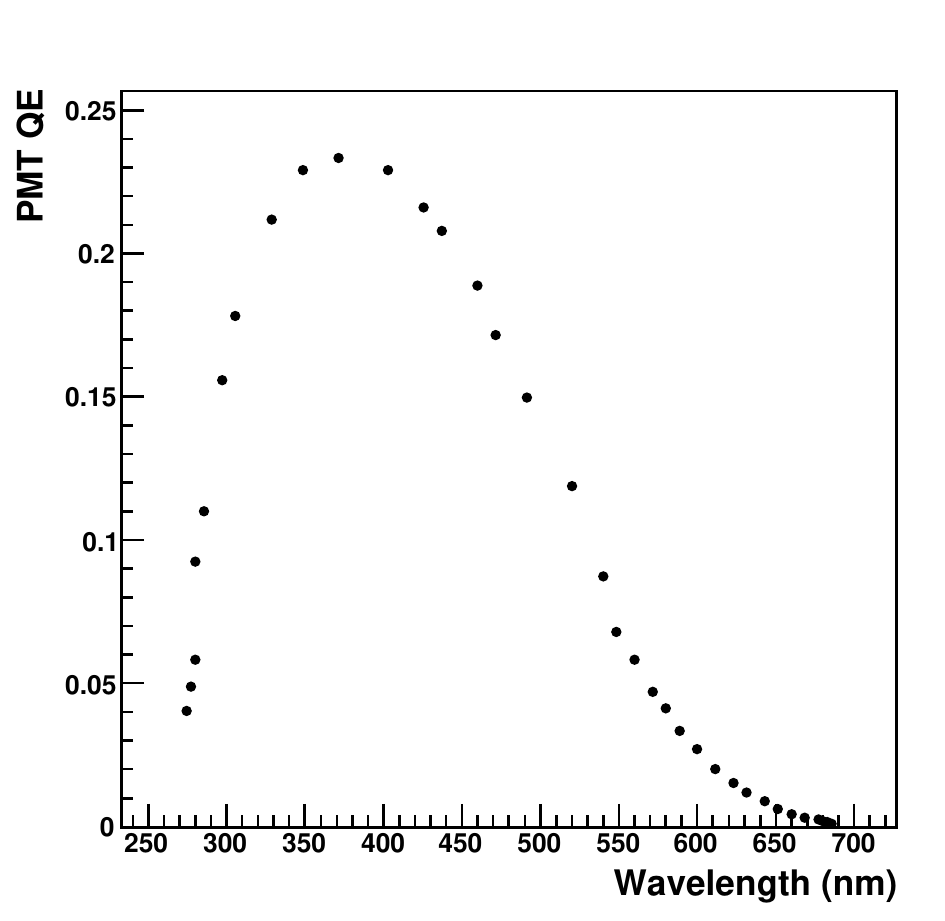} 
   \caption{PMT quantum efficiency  versus wavelength in nm for the Hamamatsu H1161 PMT~\cite{hamam_photot_h1161} used in the experimental apparatus.}
   \label{fig:qe}
\end{figure}

The PMT response is also wavelength dependent; Fig.~\ref{fig:qe} shows the quantum efficiency of the photomultiplier tube~\cite{hamam_photot_h1161}.

\subsection{CF$_4$ Spectrum}

\begin{figure}[htbp] 
   \centering
   \includegraphics[height=0.8\linewidth]{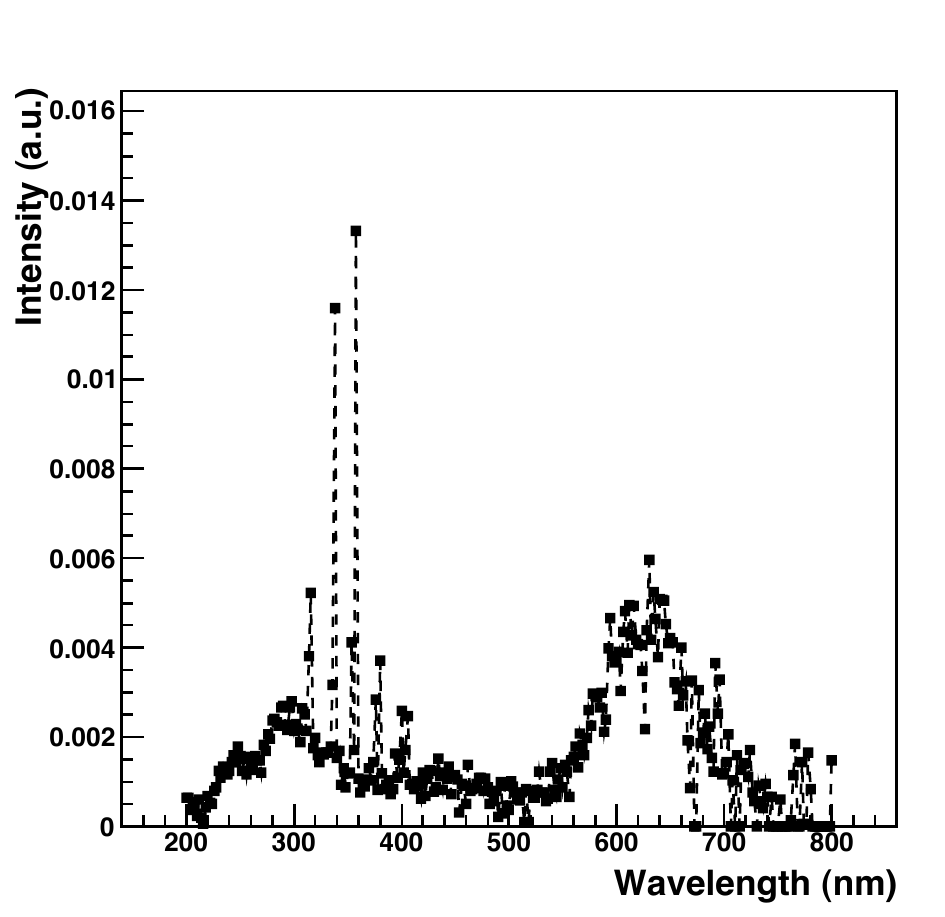} 
   \caption{Measured CF$_4$ scintillation spectrum in arbitrary units of intensity versus wavelength in nm. The integral is normalized to unity. For clarity, error bars are not shown. The spectrum is corrected for the spectrometer PMT quantum efficiency.}
   \label{fig:spectrum}
\end{figure}

To report the ratio of photons to electrons in an apparatus independent way, the observed number of photons must be corrected for the fraction of the total spectrum that is observable in this particular apparatus. Note that the spectrometer replaces the PMT in Fig~\ref{fig:box}; that is, the spectrum is measured through the same slit used in counting photons from the avalanche. Fig.~\ref{fig:spectrum} shows the measured emission spectrum. The spectrum is normalized to unit area, because the fraction of observable photons depends only on the shape of the spectrum, not its normalization. The spectrum was taken in 2~nm steps with the same spectrometer used in the transmission measurements. The spectrum has been corrected for second order diffraction at wavelengths above 400~nm; below 200~nm, oxygen in the air absorbs any photons, and so light from 200 to 400~nm has no second order component. The light intensity above 400~nm is given by the equation \[I^{obs}(\lambda) = QE(\lambda)\cdot I^{true}(\lambda) + f\cdot QE(\frac{\lambda}{2}) \cdot I^{true}(\frac{\lambda}{2})\], where $f$ is the fraction of second order light, and $QE(\lambda)$ is the wavelength dependent quantum efficiency of the spectrometer phototube. Since the quantum efficiency is known, if $f$ is known, then the above equation can be solved for $I^{true}(\lambda)$. The value of $f$ can be measured by taking the spectrum of an ultraviolet source both in the UV and in the visible, with and without a filter with a cutoff near 400~nm, and then comparing the intensity in the visible to that in the UV, after correcting for the transmission of the filter. In this spectrometer, $f = 0.25\pm0.1$.

This is the first measurement of this spectrum between 200 and 800~nm. Note that 58\% $\pm$ 6\% of the spectrum is above 450~nm. This is particularly interesting, since the quantum efficiency of most CCDs turns on around 450~nm and peaks around 600~nm, indicating that CCD readout is well-matched to the CF$_4$ spectrum. See Ref.~\cite{fraga2003gsh} for an explanation of the excitations of the CF$_4$ molecule that give rise to the 300~nm and 630~nm peaks. Note also a few transition lines in the spectrum; the origins of these lines are not described in the literature, and could be a point of further study.

\begin{figure}[htbp] 
   \centering
   \includegraphics[height=0.8\linewidth]{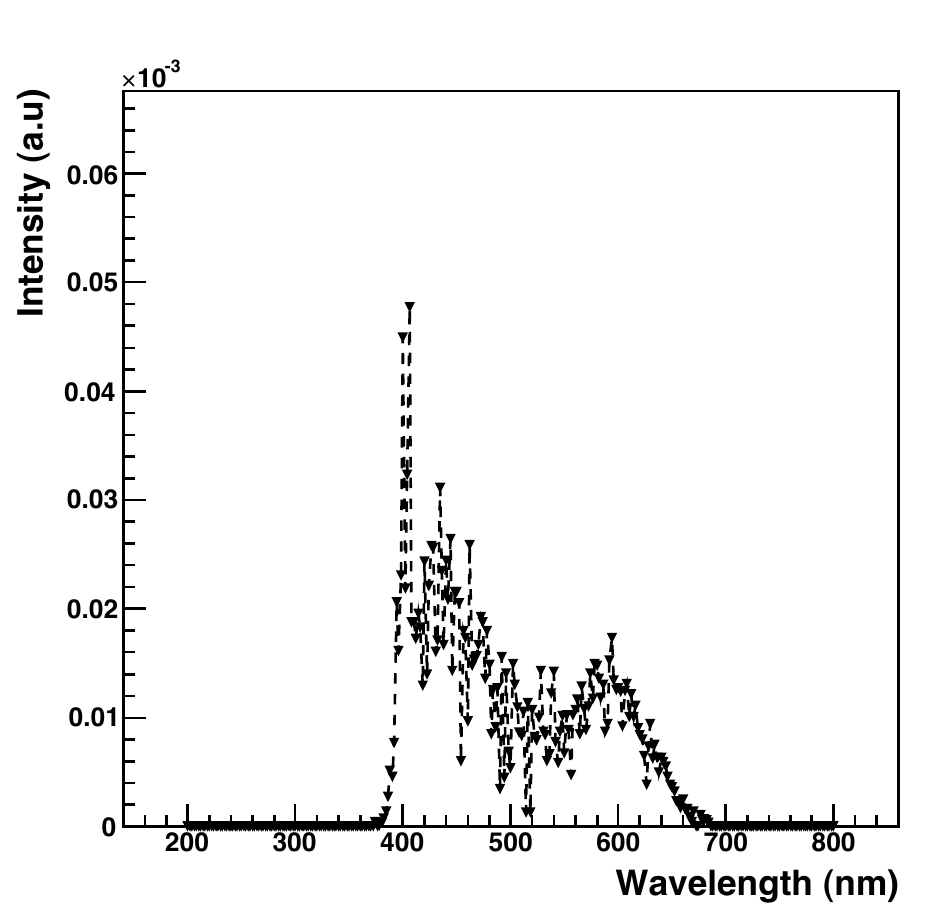} 
   \caption{Observed photon spectrum at the apparatus PMT in arbitrary units of intensity versus wavelength in nm. The scale of the intensity is the same as in Fig.~\ref{fig:spectrum}. For clarity, error bars are not shown. Each bin is the product of the true CF$_4$ spectrum, the acrylic and quartz transmittance, and the PMT quantum efficiency. }
   \label{fig:wavedep}
\end{figure}

The spectrum measured by the PMT in the proportional tube apparatus, shown in Fig.~\ref{fig:wavedep}, is computed within each wavelength bin with the formula \[I_{observed} = I_{spectrum}\cdot T_{acrylic}\cdot T_{quartz}\cdot QE_{PMT}\] where $I_{spectrum}$ is the total CF$_4$ spectrum, $T_{acrylic}$ is the transmission through acrylic, $T_{quartz}$ is the transmission through quartz, and $QE_{PMT}$ the PMT quantum efficiency. By integrating this spectrum and comparing with the raw CF$_4$ spectrum, an overall wavelength dependent correction factor of $0.0035\pm 0.0002$ is found. Figs.~\ref{fig:spectrum} and~\ref{fig:wavedep} do not show error bars, but the statistical error on the spectrum, as well as the errors contributed from background subtraction and multiplying by the quartz and acrylic transmission curves are included in finding the correction factor error. Typical relative errors on the raw spectrum are 30\% and on the convoluted spectrum are 50\%. Because the integral is performed as a sum, the relative error on the final correction factor is significantly reduced, because of the properties of propagation of errors in addition. Table~\ref{tab:corr} shows the relative size of the corrections to the photon measurement. Combining all of these corrections gives \[N^{measured}_{\gamma} = C_{sa} \cdot C_{\lambda} \cdot N_{\gamma}^{true}\] where $C_{\lambda}$ is the wavelength dependent correction factor, $C_{sa}$ is the solid angle correction factor.

\begin{table}[htbp]
   \caption{Corrections to PMT measurements. Note that while the total wavelength dependent correction is dependent on the PMT quantum efficiency, as well as the quartz and lucite transmission, it is not the direct product of the three individually integrated over the whole spectrum, but rather the product of the three in each wavelength bin, then integrated over the whole spectrum.}
   \centering
   \begin{tabular}{|l|c|c|} 
   \hline
    \hspace{1.2 cm} Correction&Value&Relative Error\\
    \hline
    PMT Calibration&1& 0.04\\
    \hline
    Solid Angle&0.00083&0.20\\
    \hline
    Total Wavelength Dependent&0.0035&0.0002\\
    \hline
     \hspace{0.3 cm} PMT Quantum Efficiency&0.064&0.002\\
    \hspace{0.3 cm} Lucite Transmission&0.71&0.04\\
    \hspace{0.3 cm} Quartz Transmission&0.124&0.007\\
    \hline
   \end{tabular}

   \label{tab:corr}
\end{table}

\section{Results}

\subsection{Voltage and Pressure Dependancies} 
\begin{figure}[htbp]
\centering
\subfigure[]{\label{fig:elphvolt}\includegraphics[width=0.8\linewidth]{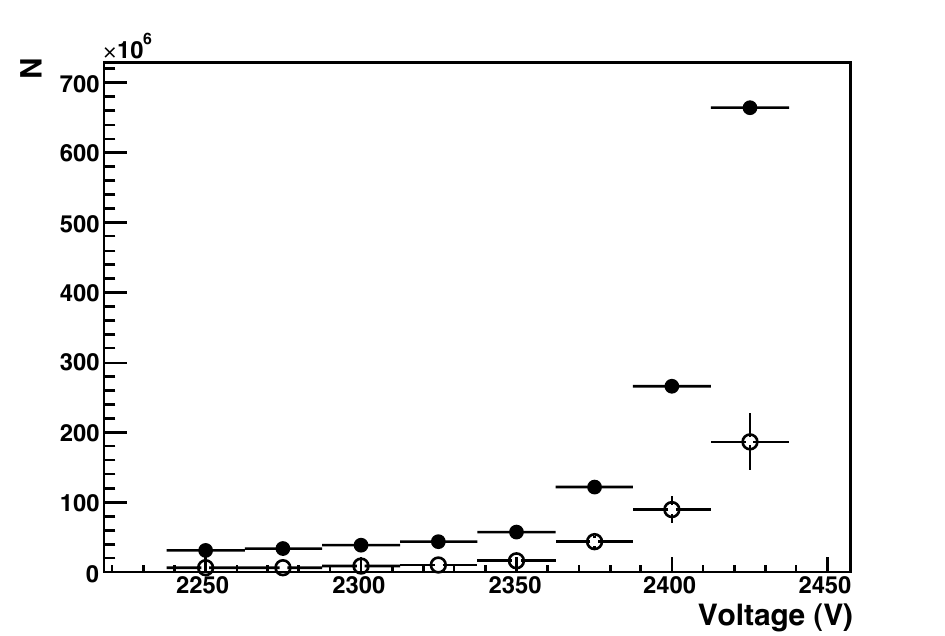}}
\subfigure[]{\label{fig:elphpress}\includegraphics[width=0.8\linewidth]{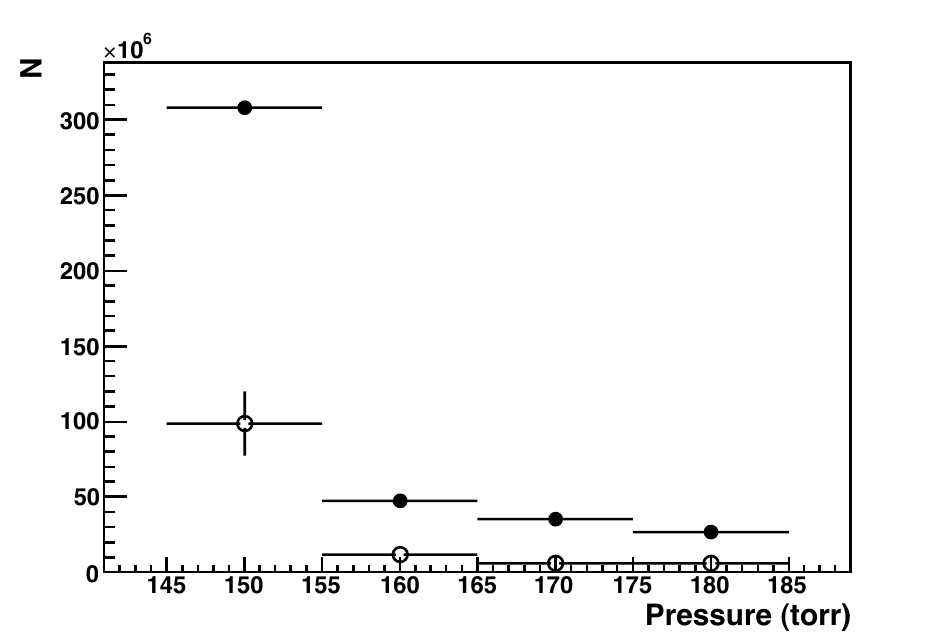}}
\caption{Pressure and voltage dependence of photon and electron production. Open circles show photons, filled circles show electrons. (a): Number of electrons  and photons versus voltage at 180 Torr CF$_4$ pressure.  (b): Number of electrons and photons versus pressure at 2325 V wire voltage.}
\label{fig:elph}
\end{figure}

The number of electrons and photons in the avalanche were measured for a range of pressures and wire voltages. Fig.~\ref{fig:elphvolt} shows the number of electrons (solid circles) and photons (open circles) produced in the avalanche as a function of voltage at 180~Torr; Fig.~\ref{fig:elphpress} shows the same as a function of pressure for 2325~V. Comparing the two processes, it is clear that the photon emission and the  electrons production are linked, as they show similar functional dependance on the two variables of wire voltage and pressure. 

The interpretation of the data, beyond the general trend that the avalanche multiplication increases sharply with voltage and decreases sharply with pressure, is difficult. CF$_4$ scintillates in ultraviolet, and the UV photons striking the side of the proportional tube eject electrons, which create secondary avalanches; as a result one Fe-55 decay can produce up to 10 avalanches. Since the number of avalanches is not constant for a given wire voltage, this makes it difficult to calculate the gain, $N^{avalanche}_{e^-}/ N^{primary}_{e^-}$. An interesting side-effect of this phenomenon is an easy way to estimate the drift velocity in CF$_4$. Since the pulses come an average of 105 ns apart, and the radius of the tube is half an inch, the corresponding drift velocity is approximately 12 cm/$\mu$s. This is in good agreement with~\cite{vavra1993med}, at the field near the edge of the tube, 35 kV/m, at 180 Torr (1.48 kV/cm/atm).

\subsection{$N_{\gamma}/N_{e^-}$}

\begin{figure}[htbp] 
   \centering
   \includegraphics[width=0.8\linewidth]{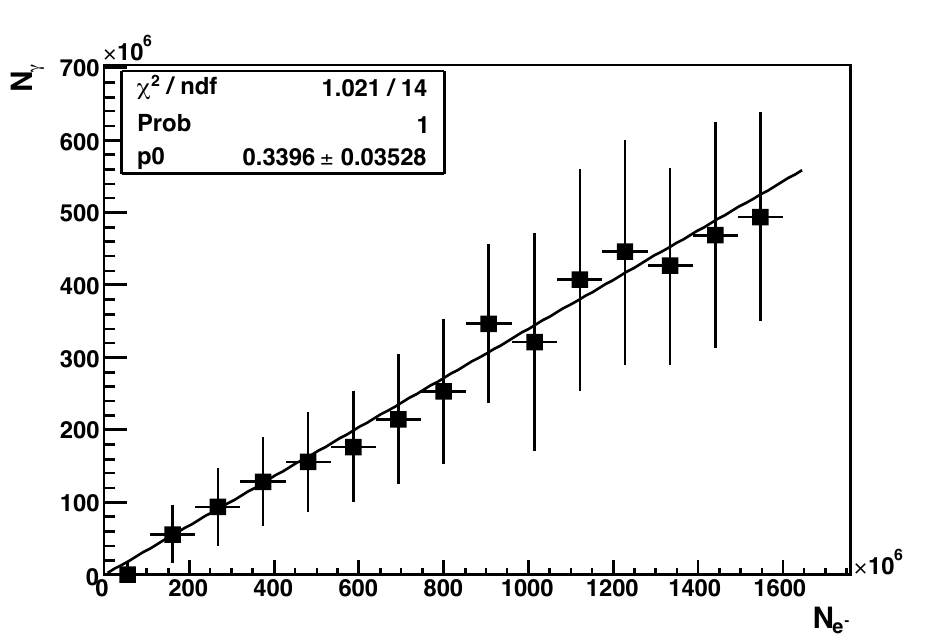} 
   \caption{Number of photons vs number of electrons, fit with a line passing through zero. The data are a combination of several data sets of varying pressure and voltage. The slope of the fit line, 0.34 $\pm$ 0.04, is the ratio of photons to electrons produced in avalanches in CF$_4$.}
   \label{fig:moneyplot}
\end{figure}

While the gain of the gas is expected to vary with pressure and voltage, the ratio of photons to electrons should be invariant, as it is   an intrinsic property of the gas. The data validates this hypothesis, and therefore all of the data can be combined to calculate the ratio. Fig.~\ref{fig:moneyplot} shows the number of photons as a function of the number of electrons for all data after all calibrations and corrections have been applied. A first-degree polynomial describes the data well; this is the functional form used to fit the data. The fitted line is also constrained to pass through the origin, corresponding to a detector limit of zero photons emitted when zero electrons are present. The slope of this line, $0.34\pm 0.04$, gives the number of photons emitted per electron in the avalanche. This ratio is measured over the range of 140-180~Torr and 2150-2425~V.

\subsection{Discussion}

The largest source of uncertainty in this measurement comes from the variations of experimental conditions between runs. This variation can come from a number of factors, the most prominent of which are variations in gas pressure and contamination. The measured leak rate of the vacuum vessel is 5x10$^{-6}$~Torr/L$\cdot$s, with the result that taking data over the span of four hours leads to contamination of 0.2\% of the CF$_4$ with air. Furthermore, different levels of contamination before filling the vessel with CF$_4$ can affect the run to run performance. Local temperature and humidity variations can also affect the level and nature of contamination, as well as the performance of the PMT. 

The error resulting from the variation between runs is taken into account in the error bars of Fig.~\ref{fig:moneyplot}. A scatterplot of $N_{\gamma}$ vs. $N_{e^-}$ is made from all the data. Then a profile plot is created using the spread of the data points in each bin of the histogram when calculating the errors, instead of the root mean square (RMS). The reason for this choice is that in each bin of number of electrons, several data sets are combined. If only one data set were used to make this plot, the projection of this bin onto the y-axis (number of photons) would be gaussian, and the error on the mean would be the appropriate error. However, since the projection is, in fact, non-gaussian, the spread better describes the error on the mean. Typical errors are 50\%. The errors introduced from the solid angle and wavelength-dependent factors are then also included in the error bars of Fig.~\ref{fig:moneyplot}, added in quadrature with the error from the spread. Note that the correct choice of error when making the plot has a significant effect on the error of the fit; using the spread instead of the RMS increases the error on the photon to electron ratio by approximately a factor of two.

The measured ratio of photons to electrons, 0.34 $\pm$ 0.04 is in agreement with the Pansky et. al. measurement of 0.3 $\pm$ 0.15~\cite{pansky1995sca}, but improves on the relative error by a factor of five. Furthermore, the Pansky measurement was made at 10~Torr, and this measurement was made at 140-180~Torr, which is more typical of current and proposed experiments using CF$_4$ as a detection technique for dark matter. 

\section{Conclusions}
This paper presents measurements of two critical properties of CF$_4$ gas: the ratio of photons to electrons produced in an avalanche, 0.34 $\pm$ 0.04, and the scintillation spectrum between 200 and 800 nm, shown in Fig.~\ref{fig:spectrum} with 10\% bin-to-bin errors, with the interesting observation that CCD quantum efficiency and the CF$_4$ spectrum are well-matched. With a firm grasp of these important properties, the prospects for optical readout of dark matter detectors using CF$_4$ as a target are good.

\section{Acknowledgments}
The authors would like to thank Ulrich Becker, Scott Sewell, and the MIT Junior Laboratory for their assistance. This work is supported by the Department of Defense National Defense Science and Engineering Graduate Fellowship, the Pappalardo Fellowship, and the Department of Energy Advanced Detector Research Program.

\bibliography{CF4_writeup}
\bibliographystyle{elsart-num}

\pagebreak

\appendix
\section{Table of Values for CF$_4$ Spectrum}

Note that where the intensity given is 0, this is due to background subtraction (the fluctuation dominates in that bin), but in all cases where the intensity is ``negative'', the value is consistent with zero.

\begin{table}[htbp]
     \caption{CF$_4$ spectrum intensity, from 200-348~ nm, unnormalized.}
\begin{tabular}{cccccccccccc}
\hline 
$\lambda$ (nm) & I (a.u.) & $\Delta$I (a.u.)&$\lambda$ (nm) & I (a.u.) & $\Delta$I (a.u.)&$\lambda$ (nm) & I (a.u.) & $\Delta$I (a.u.)\\
\hline
200&65.3145&83.1793&250&159.993&52.9829&300&217.379&56.7517\\
202&64.9513&77.1907&252&126.838&51.6025&302&233.081&57.3984\\
204&37.7948&70.909&254&154.446&52.5492&304&217.642&56.9136\\
206&54.3224&67.1186&256&119.075&51.0791&306&190.076&55.9737\\
208&40.0184&62.5843&258&156.888&52.4203&308&267.131&58.8478\\
210&23.9843&58.6422&260&134.347&51.4503&310&252.926&58.4195\\
212&61.476&58.664&262&128.501&51.1157&312&216.498&57.1792\\
214&20.384&55.4873&264&159.68&52.3001&314&384.908&63.1203\\
216&5.66265&53.4733&266&149.039&52.0261&316&528.661&67.8059\\
218&43.0572&53.7333&268&148.662&52.1418&318&176.739&55.9375\\
220&68.6155&53.5294&270&122.797&51.2778&320&200.179&56.8896\\
222&52.9467&51.6911&272&185.443&53.7872&322&160.849&55.4929\\
224&50.8944&50.4654&274&171.15&53.3864&324&146.318&55.016\\
226&80.9932&51.2871&276&208.986&54.9209&326&166.361&55.8361\\
228&88.2004&51.4579&278&199.511&54.7078&328&163.823&55.74\\
230&125.335&52.8039&280&239.663&56.3021&330&166.361&55.8361\\
232&110.288&52.0986&282&244.4&56.6083&332&167.207&55.8682\\
234&136.568&53.0044&284&236.57&56.4647&334&180.317&56.3622\\
236&119.53&52.2277&286&229.103&56.3333&336&321.574&61.4338\\
238&117.207&52.0215&288&269.804&57.9256&338&1175.04&85.9279\\
240&126.249&52.2585&290&273.044&58.1799&340&155.365&55.4181\\
242&144.956&52.8646&292&230.276&56.7941&342&170.59&55.9961\\
244&161.15&53.3663&294&218.17&56.4953&344&94.8868&53.0603\\
246&145.063&52.641&296&270.431&58.5051&346&132.104&54.5233\\
248&180.456&53.8639&298&283.269&59.0329&348&88.12&52.7899\\

\end{tabular}

   \label{tab:cf4a}
\end{table}

\begin{table}[htbp]
     \caption{CF$_4$ spectrum intensity, from 350-498~ nm, unnormalized.}
\begin{tabular}{cccccccccccc}
\hline 
$\lambda$ (nm) & I (a.u.) & $\Delta$I (a.u.)&$\lambda$ (nm) & I (a.u.) & $\Delta$I (a.u.)&$\lambda$ (nm) & I (a.u.) & $\Delta$I (a.u.)\\
\hline
350&120.262&54.0621&400&261.469&64.734&450&116.828&71.0576\\
352&121.954&54.1282&402&121.493&59.7523&452&106.724&71.2801\\
354&417.578&64.6539&404&173.932&61.8298&454&31.1921&69.6012\\
356&173.974&56.1237&406&249.323&64.9429&456&95.7981&73.3892\\
358&1349.7&90.1239&408&93.9101&59.6254&458&92.9047&73.4476\\
360&108.202&53.6816&410&96.5054&59.9044&460&51.6535&72.056\\
362&76.9956&52.5315&412&83.8492&59.6386&462&140.349&76.5034\\
364&92.0338&53.2321&414&102.526&60.6747&464&81.0909&75.5488\\
366&91.3611&53.302&416&91.7378&60.5796&466&85.2531&75.8922\\
368&106.486&54.0005&418&62.9991&60.2199&468&87.1913&76.1723\\
370&131.94&55.0955&420&123.008&63.3731&470&95.8144&77.0681\\
372&103.904&54.0939&422&69.5993&61.3572&472&111.003&78.2183\\
374&145.346&55.8105&424&108.868&62.9983&474&110.272&79.0595\\
376&287.237&61.1277&426&128.814&64.119&476&80.4428&78.8334\\
378&82.9566&53.5532&428&125.327&64.3539&478&108.772&81.0608\\
380&375.662&64.3694&430&79.1971&63.3466&480&88.0652&81.5078\\
382&119.256&55.2093&432&87.634&64.5492&482&51.5956&81.339\\
384&84.5598&54.0615&434&153.941&67.8935&484&78.4339&83.5651\\
386&95.4375&54.73&436&113.924&67.2079&486&56.1069&82.9831\\
388&104.22&55.3144&438&82.3555&66.2183&488&80.3494&84.1533\\
390&72.0601&54.2596&440&121.29&67.9524&490&21.5875&83.8431\\
392&84.3609&54.9963&442&107.651&67.8507&492&99.5699&88.7153\\
394&164.453&58.3762&444&136.337&69.3701&494&29.1645&86.2423\\
396&109.309&56.4777&446&73.7305&67.856&496&92.6976&88.7257\\
398&151.173&58.3587&448&111.892&70.2786&498&46.4752&86.7559\\

\end{tabular}

   \label{tab:cf4b}
\end{table}

\begin{table}[htbp]
     \caption{CF$_4$ spectrum intensity, from 500-648~ nm, unnormalized.}
\begin{tabular}{cccccccccccc}
\hline 
$\lambda$ (nm) & I (a.u.) & $\Delta$I (a.u.)& $\lambda$ (nm) & I (a.u.) & $\Delta$I (a.u.)&$\lambda$ (nm) & I (a.u.) & $\Delta$I (a.u.)\\
\hline
500&36.4849&86.1801&550&89.2489&114.22&600&375.581&160.983\\
502&102.641&90.3827&552&140.487&116.652&602&396.298&162.189\\
504&89.6291&91.6857&554&123.371&119.097&604&307.501&159.743\\
506&76.185&90.9067&556&66.41&120.347&606&441.883&164.152\\
508&61.0873&90.1295&558&150.069&124.65&608&487.941&165.661\\
510&59.6232&92.1965&560&156.539&126.316&610&394.243&168.623\\
512&76.4439&95.0953&562&180.365&127.912&612&502.845&178.548\\
514&9.59214&92.437&564&134.692&127.205&614&432.149&176.636\\
516&83.9446&95.1063&566&211.043&130.74&616&499.523&179.178\\
518&9.5425&92.7015&568&184.317&130.785&618&422.519&175.592\\
520&81.109&95.8827&570&161.737&133.686&620&413.263&174.281\\
522&64.99&97.0012&572&200.037&138.831&622&411.966&187.527\\
524&65.5319&98.8485&574&263.806&142.617&624&352.283&201.435\\
526&81.8819&99.7023&576&228.699&143.209&626&220.057&213.013\\
528&124.516&101.538&578&300.985&144.578&628&445.513&235.716\\
530&77.605&100.46&580&295.266&143.512&630&603.457&208.065\\
532&76.8365&101.096&582&293.906&144.389&632&422.403&178.175\\
534&56.2859&100.054&584&269.862&144.528&634&532.268&184.801\\
536&63.7774&100.12&586&303.549&150.368&636&469.454&185.658\\
538&123.261&105.324&588&215.322&152.109&638&384.704&181.83\\
540&144.125&109.465&590&242.004&155.174&640&514.759&185.582\\
542&83.1506&107.608&592&403.651&162.596&642&513.66&186.306\\
544&66.2024&107.339&594&472.805&162.011&644&511.235&187.021\\
546&105.547&111.419&596&386.357&156.666&646&458.367&187.92\\
548&131.109&115.109&598&371.513&158.476&648&415.111&189.259\\
\end{tabular}

   \label{tab:cf4c}
\end{table}

\begin{table}[htbp]
     \caption{CF$_4$ spectrum intensity, from 650-798~ nm, unnormalized.}
\begin{tabular}{cccccccccccc}
\hline 
$\lambda$ (nm) & I (a.u.) & $\Delta$I (a.u.)&$\lambda$ (nm) & I (a.u.) & $\Delta$I (a.u.)&$\lambda$ (nm) & I (a.u.) & $\Delta$I (a.u.)\\
\hline
650&426.926&191.24&700&120.682&225.586&750&0&293.755\\
652&417.266&192.497&702&146.175&256.884&752&60.5802&287.798\\
654&327.236&191.243&704&207.998&300.511&754&0&316.024\\
656&311.523&192.665&706&0&262.753&756&0&362.564\\
658&273.808&193.221&708&105.268&244.81&758&0&324.491\\
660&406.299&199.916&710&0&413.9&760&0&304.701\\
662&297.686&198.718&712&161.009&663.26&762&14.0033&310.697\\
664&329.468&202.539&714&0&403.65&764&115.306&316.502\\
666&194.843&210.949&716&127.787&242.856&766&186.199&325.06\\
668&86.9937&223.037&718&140.813&243.233&768&0&318.339\\
670&330.508&342.312&720&144.438&243.434&770&0&324.114\\
672&0&473.582&722&111.954&245.155&772&145.036&341.582\\
674&0&312.183&724&172.941&250.655&774&6.19413&338.65\\
676&309.099&209.622&726&76.1387&248.95&776&4.21727&341.51\\
678&189.65&208.318&728&57.3759&250.348&778&167.01&354.912\\
680&212.602&212.348&730&0&250.834&780&85.5268&358.783\\
682&256.281&210.695&732&93.4533&258.485&782&0&367.785\\
684&175.286&205.101&734&56.6552&261.381&784&0&380.855\\
686&225.424&211.389&736&42.0142&265.915&786&0&380.178\\
688&156.446&213.468&738&95.5614&268.966&788&0&382.555\\
690&123.705&211.65&740&67.5604&269.02&790&0&398.214\\
692&369.487&220.837&742&0&273.198&792&0&405.17\\
694&255.217&220.764&744&67.177&283.406&794&0&417.16\\
696&332.219&228.197&746&0&298.388&796&0&450.477\\
698&118.388&222.757&748&0&310.516&798&0&445.163\\
\end{tabular}

   \label{tab:cf4d}
\end{table}

\end{document}